\documentclass[
    ,final            
  ]
  {aipproc}

\layoutstyle{6x9}

\begin{document}

\title{Cluster magnetic fields from active galactic nuclei}


\author{P.M.~Sutter}{
  address={Department of Physics, 
           University of Illinois at Urbana-Champaign, 
           Urbana, IL 61801-3080}
}

\author{P.M.~Ricker}{
  address={Department of Astronomy, 
           University of Illinois at Urbana-Champaign, 
           Urbana, IL 61801}
  ,altaddress={National Center for Supercomputing Applications, 
               University of Illinois at Urbana-Champaign, 
               Urbana, IL 61801}
}

\author{H.-Y.~Yang}{
  address={Department of Astronomy, 
           University of Illinois at Urbana-Champaign, 
           Urbana, IL 61801}
}

\begin{abstract}
Active galactic nuclei (AGN) found at the centers of clusters of
galaxies are a possible source for weak cluster-wide magnetic
fields. To evaluate this scenario, we present 3D adaptive mesh refinement
MHD simulations of a cool-core cluster that include injection of
kinetic, thermal, and magnetic energy via an AGN-powered jet. 
Using the MHD solver in FLASH 2, 
we compare several sub-resolution approaches 
that link the estimated accretion rate as measured on
the simulation mesh to the accretion rate onto the central black hole
and the resulting feedback. We examine the effects 
of magnetized outflows on the accretion history of the black hole
and discuss the ability of these models to magnetize the cluster medium.
\end{abstract}

\maketitle


\section{Introduction}

Faraday rotation measures of galaxy clusters suggest they contain 
magnetic fields with average strengths of $\sim 1$~$\mu$G 
(e.g., \citep{Feretti}), 
but the origin of these fields is unknown. Possible seed fields, 
such as those generated by the Biermann battery mechanism, 
require amplification \citep{Widrow}.
Accretion disks around active galactic nuclei (AGN) 
are a promising candidate amplification source, 
since AGN-launched jets require 
magnetic fields and the energetics of AGN-blown bubbles 
suggest that AGN can amplify the seed fields to the observed strengths. 

We would like to directly test this scenario in simulation, 
but since AGN accretion disks are typically smaller then $\sim 100$ AU, 
and cosmological 
simulations have resolutions on the order of $\sim 1$ kpc, we must 
include the amplification and jet launching as a subgrid model.
We investigate two commonly-used AGN feedback models: 
one based on jets \citep{Cattaneo}, 
the other on already-inflated bubbles \cite{Sijacki}.
We link the feedback energies of 
these models to the strength of an injected magnetic field. 
We inject magnetic fields with the form described by \cite{Li}, in which 
the field has both toroidal and poloidal components.

\section{Analytical and Numerical Approach}

We assume Bondi accretion:
\begin{equation}
  \dot{M}_{\rm Bondi} = 4 \pi G^2 m^2_{\rm BH} \rho / {c_{\rm s}}^3,
  \label{eq:accretionBondi}
\end{equation}
where the sound speed $c_{\rm s}$ and the density $\rho$ are measured 
on the simulation mesh, and $m_{\rm BH}$ is the black hole mass.
Following \cite{Sijacki}, to compensate for 
under-resolving the actual accretion disk, 
we assume a constant multiple of the Bondi rate:
\begin{equation}
  \dot{M} = \alpha \dot{M}_{\rm Bondi}
  \label{eq:accretion}
\end{equation}
with $\alpha=100$.

We follow \cite{Cattaneo} for modeling jet-based feedback  
where the energy injection rate is
\begin{equation}
  \dot{E} = \epsilon_{\rm F} \dot{M} c^2 
            \left( 1-1/M_{\rm load} \right) | \Psi |.
  \label{eq:feedbackJetEner}
\end{equation}
Similarly, the momentum injection rate is
$\dot{\bf{P}} = \sqrt{2 \epsilon_{\rm F}} \dot{M} c \Psi$ ,
and the mass injection rate is
$\dot{M}_{\rm inj} = M_{\rm load} \dot{M} | \Psi |$.

The window function $\Psi$, 
which provides a mapping onto the mesh, is
\begin{equation}
  \Psi({\bf x}) = \frac{1}{2 \pi r_{\rm ej}^2} 
         \exp{\left(-\frac{x^2+y^2}{2 r_{\rm ej}^2}\right)}  
         \frac{z}{h_{\rm ej}^2}.
  \label{eq:psi}
\end{equation}
We cut off injection at $z=h_{\rm ej}$ and $r=2.6 r_{\rm ej}$.
In the above, $c$ is the speed of light, 
and we will define the injection region through 
$r_{\rm ej}=3.2$ kpc and $h_{\rm ej}=2.5$ kpc. The injection region is oriented 
along the z-axis. We assume a jet 
mass loading factor of $M_{\rm load}=100$ and feedback efficiency of 
$\epsilon_{\rm F}=0.1$.

For bubble injection, as in \cite{Sijacki}, we have only thermal 
energy injection:
\begin{equation}
  \dot{E} = \epsilon_{\rm F} \epsilon_{\rm M} \dot{M} c^2,
  \label{eq:feedbackBub}
\end{equation}
where $\epsilon_{\rm F}$ is the same as above and 
$\epsilon_{\rm M}=1.0$.
We distribute this energy uniformly in a sphere with radius 
determined by
\begin{equation}
  R_{\rm bub}=R_{\rm 0} \left( \frac{\dot{E} dt}{E_{\rm 0}} 
                     \frac{\rho_{\rm 0}}{\rho} \right)^{1/5},
  \label{eq:rbub}
\end{equation}
where we define the scalings by $R_0=43$ kpc, $E_{\rm 0}=10^{60}$ erg, 
$\rho_{\rm 0}=10^6 \: {\rm M}_\odot {\rm kpc}^{-3}$, and dt is the timestep. 
These scalings ensure that 
a bubble in a typical cluster environment will have 
a realistic size.
The bubbles are always centered on the black hole, 
and we only form bubbles when the black hole 
has increased its mass since the previous bubble formation by 
$\Delta M_{\rm BH}/M_{\rm BH}  > 0.001$.

The injected magnetic field takes the form
\begin{eqnarray}
  B_r(r',z')    & = & 2 B_{\rm 0} z' r' \exp{\left( -{r'}^2 - {z'}^2 \right)} \\
  B_z(r',z')    & = & 2 B_{\rm 0} \left( 1 - {r'}^2 \right)
                 \exp{\left( -{r'}^2 - {z'}^2 \right)} \\ 
  B_\phi(r',z') & = & B_{\rm 0} \alpha_{\rm B} r' 
                 \exp{\left( -{r'}^2 - {z'}^2 \right)}, 
  \label{eq:mag}
\end{eqnarray}
where $r'=\sqrt{x^2+y^2}/r_{\rm 0}$ and $z'=z/r_{\rm 0}$.
Here, $r_0$ is $1/2 R_{\rm bub}$ for bubbles and $1/2 R_{\rm ej}$ for jets,
 and $\alpha_{\rm B}$ is the ratio 
of of polodial to toroidal flux. We choose $\alpha_{\rm B}=\sqrt{10}$ 
for an initially relaxed field, as suggested by \cite{Li}. 
We determine the scale $B_{\rm 0}$ by giving half of the available 
feedback energy to the magnetic field. 

We performed three-dimensional simulations with an isolated 
cluster profile in a 2048 kpc box using FLASH 2.5 \cite{Fryxell}, 
an adaptive mesh refinement (AMR) code. 
Both the jet and bubbles runs used a maximum resolution of 
1.0 kpc within the central 
50 kpc region.
The AGN began as a $10^7 M_\odot$ black hole in the center 
of an NFW (\cite{NFW})  
gravitational potential. The cluster had a concentration of $6.5$, scaling 
radius of $165$ kpc, 
total mass of $10^{14} \: {\rm h}^{-1}{\rm M}_\odot$, 
and gas fraction of $0.12$. 
We included cooling from \cite{Sutherland} 
assuming $1/3$ solar metallicity. We allowed the cluster 
to relax for $\sim 1$~Gyr before activating cooling and feedback.  

\section{Results}
\label{sec:Results}
\begin{figure}
  \includegraphics[height=.3\textheight]{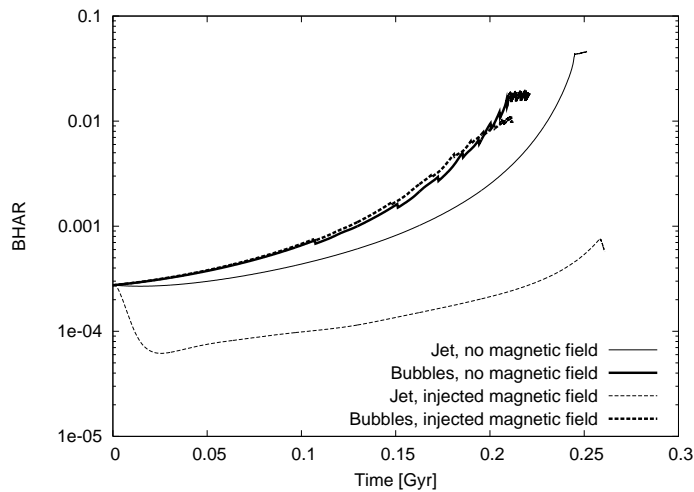}
  \caption{Accretion rate as a fraction of the Eddington rate 
           for magnetized and unmagnetized jets 
           and bubbles.} 
  \label{fig:accRate}
\end{figure}

Figure~\ref{fig:accRate} shows the accretion history of the black 
hole with both magnetized and unmagnetized jet- and bubble-based feedback
 until the cooling catastrophe occurs. We find that magnetizing bubbles 
does not greatly alter the accretion rate relative 
to unmagnetized injections, since bubbles occur infrequently. 
However, magnetizing jets greatly 
reduces the accretion rate. Here, the combination of an axial jet 
and a toroidal magnetic field prevents gas from accreting. 
\begin{figure}
  \includegraphics[height=.3\textheight]{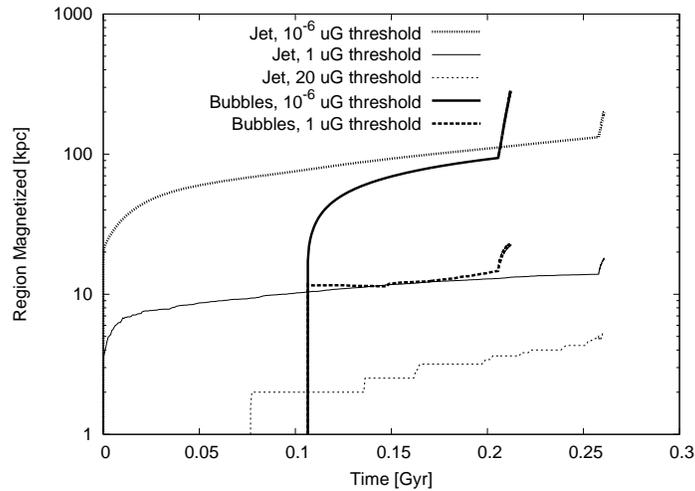}
  \caption{Magnetized volume for various thresholds. The y-axis 
           is the the cube root of the total volume that lies above
           each threshold.}
  \label{fig:magVol}
\end{figure}

We show the volume magnetized in Figure~\ref{fig:magVol}.
Both jets and bubbles are able to weakly magnetize large volumes. 
These fields may be further amplified in realistic clusters 
by turbulence and merger shocks. 
However, significant ($> 1 \: \mu {\rm G}$) fields do not penetrate 
far from the cluster core. Finally, only the jet is able to produce 
greater than $20 \: \mu {\rm G}$ fields, but these quickly dissipate. 
The outflows begin to greatly enhance the magnetization of the cluster at the 
onset of the cooling catastrophe.

With the parameter values used, we find that these feedback 
mechanisms do not provide enough heating to prevent the 
cooling catastrophe. We are investigating the new MHD solver 
in FLASH 3 (\cite{Lee}) to determine whether or not this is a 
numerical effect.

The evolution of our injected magnetic fields resembles that
of \cite{Li}, but they consider only pure magnetic injection 
from a single source with fixed energy input. Our energy input 
depends on the accretion rate, and we find much less available 
energy than the value they assume ($\sim 10^{60}$ ergs).


\begin{theacknowledgments}
The authors acknowledge support under a Presidential Early
Career Award from the U.S. Department of Energy,
Lawrence Livermore National Laboratory (contract B532720), 
a DOE Computational Science Graduate Fellowship (DE-FG02-97ER25308), 
NASA Headquarters under the NASA Earth and Space
Science Fellowship Program (NNX08AZ02H).
and the National Center for
Supercomputing Applications.
The software used in this work was in part developed by the DOE-supported ASC
/ Alliance Center for Astrophysical Thermonuclear Flashes
at the University of Chicago.
\end{theacknowledgments}

\bibliography{ms}
\bibliographystyle{aipproc}   
\nocite{*}

\end{document}